\documentclass[preprint,showpacs,amsmath,amssymb]{revtex4}
\usepackage{mathrsfs}
\usepackage{graphicx,color}
\usepackage{multirow}
\usepackage{dcolumn}
\usepackage{bm}

\begin{document}

\title{ $\beta$-decay half-lives of neutron-rich nuclei and matter flow in the $r$-process}

\author{Z. M. Niu$^{1,2}$}
\author{Y. F. Niu$^{2,3}$}
\author{H. Z. Liang$^{2,4}$}
\author{W. H. Long$^5$}
\author{T. Nik\v{s}i\'{c}$^6$}
\author{D. Vretenar$^6$}
\author{J. Meng$^{2,7,8}$}\email{mengj@pku.edu.cn}
\affiliation{$^1$School of Physics and Material Science, Anhui University,
             Hefei 230039, China}
\affiliation{$^2$State Key Laboratory of Nuclear Physics and Technology, School of Physics, Peking University,
             Beijing 100871, China}
\affiliation{$^3$Institute of Fluid Physics, China Academy of Engineering Physics,
             Mianyang 621900, China}
\affiliation{$^4$RIKEN Nishina Center,
             Wako 351-0198, Japan}
\affiliation{$^5$School of Nuclear Science and Technology, Lanzhou University,
             Lanzhou 730000, China}
\affiliation{$^6$Physics Department, Faculty of Science, University of Zagreb,
             Croatia}
\affiliation{$^7$School of Physics and Nuclear Energy Engineering, Beihang University,
             Beijing 100191, China}
\affiliation{$^8$Department of Physics, University of Stellenbosch,
             Stellenbosch 7602, South Africa}

\date{\today}

\begin{abstract}
The $\beta$-decay half-lives of neutron-rich nuclei with $20\leqslant
Z\leqslant 50$ are systematically investigated using the newly developed
fully self-consistent proton-neutron quasiparticle random phase
approximation (QRPA), based on the spherical relativistic
Hartree-Fock-Bogoliubov (RHFB) framework. Available data are reproduced
by including an isospin-dependent proton-neutron pairing interaction in
the isoscalar channel of the RHFB+QRPA model. With the calculated
$\beta$-decay half-lives of neutron-rich nuclei a remarkable speeding up
of $r$-matter flow is predicted. This leads to enhanced $r$-process
abundances of elements with $A\gtrsim 140$, an important result for the
understanding of  the origin of heavy elements in the universe.
\end{abstract}


\maketitle


Nuclear $\beta$-decay plays an important role not only in nuclear physics
but also in other branches of science, notably astrophysics and particle
physics. In nuclei the investigation of $\beta$-decay provides information
on the spin and isospin dependence of the effective nuclear interaction,
as well as on nuclear properties such as masses~\cite{Lunney2003RMP},
shapes~\cite{Nacher2004PRL}, and energy levels~\cite{Tripathi2008PRL}. In
nuclear astrophysics $\beta$-decay  half-lives set the time scale of the
rapid neutron-capture process ($r$-process), and hence determine the
production of heavy elements in the universe~\cite{Burbidge1957RMP,
Qian2007PRp, Langanke2003RMP}. In particle physics $\beta$-decay was used
to obtain the first experimental evidence of parity
violation~\cite{Wu1957PR}, and can be utilized to verify the unitarity of the
Cabibbo-Kobayashi-Maskawa (CKM) matrix~\cite{Liang2009PRC, Hardy2010RPP}.

Important advances in the measurement of nuclear $\beta$-decay half-lives
have been achieved in recent years with the development of radioactive
ion-beam facilities, especially for nuclei near the neutron shell closures
of $50$ and $82$~\cite{Audi2003NPA, Hosme2005PRL, Pereira2009PRC}. Quite
recently $\beta$-decay half-lives of very neutron-rich Kr to Tc isotopes
with neutron number between $N= 50$ and $82$ have been
measured~\cite{Nishimura2011PRL}. The new experimental results indicate a
systematic deviation from half-lives predicted by standard calculations
based on the finite-range droplet model (FRDM) plus quasiparticle random
phase approximation (QRPA)~\cite{Moller1997ADNDT}. The impact of newly
measured $\beta$-decay half-lives on the $r$-process nucleosynthesis has
been investigated in Ref.~\cite{Nishimura2012PRC}, and it has been shown
that the main effect is an enhancement in the abundances of isotopes with
mass number $A = 110 - 120$, relative to abundances calculated using
$\beta$-decay half-lives estimated with the FRDM+QRPA.

Most neutron-rich nuclei relevant for the $r$-process are still out
of experimental reach and, therefore, $r$-process
calculations are based on theoretical predictions for $\beta$-decay
 half-lives. Theoretical investigations of the nuclear $\beta$-decay
started in the 1930's with the famous Fermi
theory~\cite{Fermi1934ZP}. Two types of microscopic approaches are
nowadays mainly used in large-scale calculations of nuclear
$\beta$-decay  half-lives: the shell model and the
proton-neutron QRPA. Specifically, the shell model has been applied
to studies of $\beta$-decay  half-lives for nuclei at neutron number
$N = 50, 82, 126$, and the experimental
half-lives are reproduced by
calculations \cite{Langanke2003RMP, Pinedo1999PRL, Suzuki2012PRC}.
However, shell-model calculations for heavy nuclei away from the
magic numbers are not feasible yet because of extremely large
configuration spaces. Compared to this approach, the proton-neutron
QRPA can be applied to arbitrary heavy systems. Nuclear
$\beta$-decay calculations have been carried out using the QRPA
based on the FRDM~\cite{Moller1997ADNDT}, the extended Thomas-Fermi
plus Strutinsky integral (ETFSI) model~\cite{Borzov2000PRC}, the
Skyrme Hartree-Fock-Bogoliubov (SHFB) model~\cite{Engel1999PRC}, and
the density functional of Fayans (DF)~\cite{Borzov1996ZPA}.

Covariant density functional theory (CDFT) has been applied very
successfully to the description of a variety of nuclear structure
phenomena~\cite{Meng2006PPNP, Vretenar2005PRp, Niksic2011PPNP}. The CDFT
framework naturally includes the nucleon spin degree of freedom, and the
resulting nuclear spin-orbit potential automatically emerges with the
empirical strength, thus producing a good agreement with the experimental
single-nucleon spectrum~\cite{Liang2011PRC}. Based on the relativistic
Hartree-Bogoliubov (RHB) model in the CDFT framework, the QRPA has been
formulated~\cite{Paar2004PRC} and employed in the calculations of
$\beta$-decay half-lives of neutron-rich nuclei in the regions of
$N\approx 50$ and $N\approx 82$~\cite{Niksic2005PRC, Marketin2007PRC}.

To reliably predict properties of thousands of unknown nuclei relevant to
the $r$-process, the self-consistency of the QRPA approach is essential.
Only recently the fully self-consistent relativistic RPA has been
formulated based on the relativistic Hartree-Fock (RHF)
theory~\cite{Liang2008PRL}. The RHF+RPA model
produces results in excellent agreement
with data  on the Gamow-Teller (GTR) and spin-dipole resonances
 in doubly magic nuclei, without any readjustment of the
parameters of the covariant energy density functional~\cite{Liang2008PRL,
Liang2012PRC}. Recently also the relativistic Hartree-Fock-Bogoliubov
(RHFB) theory has been developed, thus providing a unified description of both
mean field and pairing correlations~\cite{Long2010PRC,Ebran2011PRC}.

In this Letter we report the implementation of a fully self-consistent
proton-neutron QRPA for spherical nuclei, based on the RHFB
theory~\cite{Long2010PRC,Ebran2011PRC}, and its first systematic
application in calculations of $\beta$-decay half-lives of neutron-rich
nuclei with $20\leqslant Z\leqslant 50$, extending over the whole
$r$-process path from $N=50$ to $N=82$. The effect on $r$-process
nucleosynthesis simulations is also investigated using the classical
$r$-process model.


The details of the QRPA formalism in the canonical basis can be
found in Refs.~\cite{Engel1999PRC, Paar2004PRC}. In the RHFB+QRPA
model both the direct and exchange terms are taken into account, so
the matrix elements of the particle-hole ($p$-$h$) $V^{ph}$ and
particle-particle ($p$-$p$) $V^{pp}$ residual interactions read,
respectively,
\begin{eqnarray}
      V_{pnp'n'}^{ph}
  &=& \int\int d\boldsymbol{r}_1d\boldsymbol{r}_2
       f_p^+(\boldsymbol{r}_1) f_{n'}^+(\boldsymbol{r}_2)
       \sum_{\phi_i} V_{\phi_i}(1,2) \nonumber\\
  & &  \left[ f_{p'}(\boldsymbol{r}_2) f_n(\boldsymbol{r}_1)
             -f_n(\boldsymbol{r}_2) f_{p'}(\boldsymbol{r}_1) \right],
\end{eqnarray}
\begin{eqnarray}
      V_{pnp'n'}^{pp}
  &=& \int\int d\boldsymbol{r}_1d\boldsymbol{r}_2
      f_p^+(\boldsymbol{r}_1) f_{n}^+(\boldsymbol{r}_2)
      \sum_{T=1,0} V_{T}(1,2) \nonumber\\
  & & \left[ f_{n'}(\boldsymbol{r}_2) f_{p'}(\boldsymbol{r}_1)
            -f_{p'}(\boldsymbol{r}_2) f_{n'}(\boldsymbol{r}_1) \right] \;.
\end{eqnarray}
$p, p'$, and $n, n'$ denote proton and neutron quasiparticle canonical
states, respectively, and $\phi_i$ and $T$ denote corresponding coupling
channels. $f_{p(n)}$ are canonical wave functions extracted from the RHFB
calculations.

Because of the presence of exchange terms, the proton-neutron QRPA
interaction contains terms generated not only by the isovector meson
exchange ($\rho$ and $\pi$), but also by the exchange of isoscalar mesons
($\sigma$ and $\omega$). As in Ref.~\cite{Liang2008PRL}, the pionic
zero-range counter term introduced to remove the contact part of the
pseudovector $\pi$-N coupling, reads
\begin{eqnarray}
  V_{\pi\delta}(1,2) = g'\left[\frac{f_\pi}{m_\pi}\vec\tau\gamma_0\gamma_5\boldsymbol\gamma\right]_1
                         \cdot
                         \left[\frac{f_\pi}{m_\pi}\vec\tau\gamma_0\gamma_5\boldsymbol\gamma\right]_2
                         \delta(\boldsymbol r_1-\boldsymbol r_2)\;,
\end{eqnarray}
where the self-consistency of the model requires $g'=1/3$. For the
isovector ($T=1$) $p$-$p$ channel the pairing part of the Gogny
force D1S~\cite{Berger1984NPA} is consistently used both in the RHFB
and QRPA calculations. For the isoscalar ($T=0$) proton-neutron
pairing in the QRPA we employ a similar interaction that was
previously used in Refs.~\cite{Engel1999PRC, Niksic2005PRC,
Marketin2007PRC}:
\begin{eqnarray}
    V_{T=0}(1,2)=-V_0 \sum_{j=1}^2 g_j e^{-[(\boldsymbol{r}_1-\boldsymbol{r}_2)/\mu_j]^2} \hat{\prod}_{S=1,T=0},
\end{eqnarray}
with $\mu_1=1.2$ fm, $\mu_2=0.7$ fm, $g_1=1$, $g_2=-2$.  The operator
$\hat{\prod}_{S=1,T=0}$ projects onto states with $S=1$ and $T=0$. $V_0$
is the overall strength of the $T=0$ proton-neutron pairing. Experimental
evidence for $T=0$ proton-neutron pairing is supported by recent studies
of the level structure of $^{92}$Pd~\cite{Cederwall2011Nature}, and the
high-spin isomer in $^{96}$Cd~\cite{Singh2011PRL}.

With the individual transition strengths $B_m$ obtained from QRPA
calculations, the $\beta$-decay half-life of an even-even nucleus is
calculated in the allowed Gamow-Teller approximation using the expression:
\begin{eqnarray}\label{Eq:BetaDecayRate}
    T_{1/2}
  =\frac{D}
        {g_A^2
         \sum_m B_m f(Z,E_m)},
\end{eqnarray}
where $D=6163.4\pm3.8$ s and $g_A=1$.
The sum  runs over all
final states with an excitation energy smaller than the $Q_\beta$
value. The integrated phase volume
$f(Z,E_m)$ can be written as
\begin{eqnarray}
  f(Z,E_m) = 
               \int_{m_e}^{E_m}
               p_e E_e (E_m-E_e)^2 F_0(Z,E_e)dE_e,
\end{eqnarray}
where $p_e$, $E_e$, and $F_0(Z,E_e)$ denote the momentum,
energy, and Fermi function of the emitted electron, respectively~\cite{Langanke2003RMP}.
The $\beta$-decay transition energy $E_m$, that is, the energy difference
between the initial and final state, is calculated using the QRPA:
\begin{equation}\label{Eq:BetaDecayEm}
    E_m = \Delta_{np} - E_{\textrm{QRPA}},
\end{equation}
where $E_{\textrm{QRPA}}$ is the QRPA energy with respect to the
ground-state of the parent nucleus and corrected by the difference of the
neutron and proton Fermi energies in the parent nucleus
\cite{Engel1999PRC}, and $\Delta_{np}$ is the neutron-proton mass
difference.


\begin{figure}
\includegraphics[width=10cm]{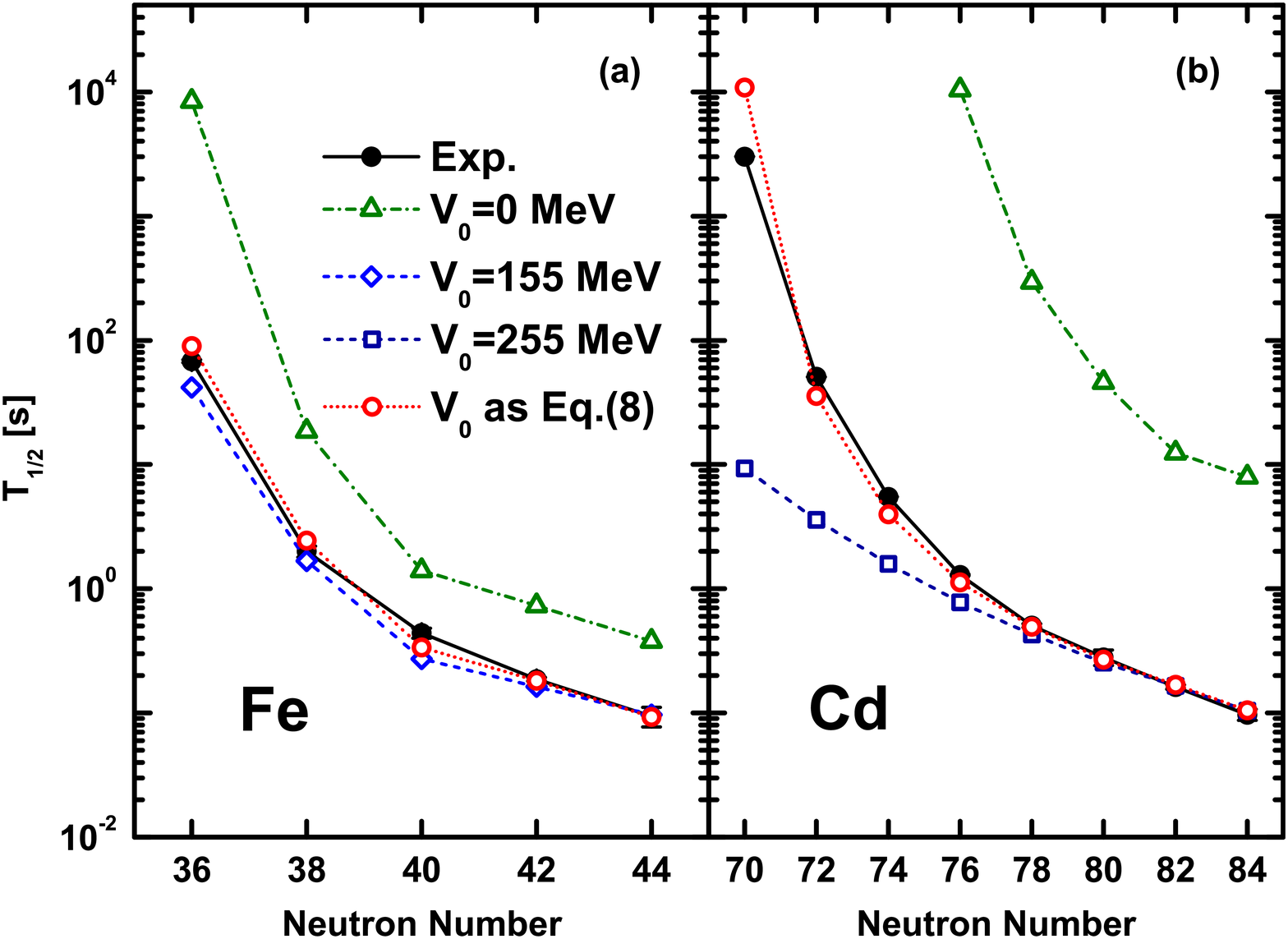}
\caption{(Color online) Nuclear $\beta$-decay half-lives of Fe (left)
and Cd (right) isotopes, calculated with the PKO1
effective interaction~\cite{Long2006PLB}, compared to the
experimental values~\cite{Audi2003NPA}. Open triangles, diamonds, and
squares denote values obtained using the RHFB+QRPA with the strength
parameter of the $T=0$ pairing: $V_0=0, 155$, and $255$ MeV,
respectively. The RHFB+QRPA values obtained with the $V_0$ of
Eq.~(\ref{Eq:V0NSubZ}) are denoted by open circles.} \label{fig1}
\end{figure}

Nuclear $\beta$-decay half-lives are very sensitive to the $T=0$
proton-neutron pairing interaction, and its strength $V_0$ is
determined by adjusting QRPA results to empirical
half-lives~\cite{Engel1999PRC, Niksic2005PRC, Marketin2007PRC}.
Taking $^{70}$Fe and $^{130}$Cd as reference nuclei for the two mass
regions, the value of $V_0$ is determined as $155$ MeV and $255$
MeV, respectively. Using these two values, the calculated half-lives
of Fe and Cd isotopes are shown in Fig.~\ref{fig1}. For comparison,
the experimental values and the results of a calculation without the
$T=0$ pairing are also displayed. One notices that the $\beta$-decay
half-lives calculated without the inclusion of $T=0$ pairing are
systematically much longer than the experimental half-lives, both
for Fe and Cd isotopes. In previous studies a constant value of
$V_0$ was taken for one isotopic chain or one mass region,
determined by adjusting to the known half-lives of selected nuclei
in the isotopic chain~\cite{Niksic2005PRC}, or several nuclei in the
same mass region~\cite{Engel1999PRC}. This procedure, of course,
limits the predictive power of the model. Moreover, as shown in
Fig.~\ref{fig1}, when $V_0$ is determined by the
$\beta$-decay half-life of $^{130}$Cd the calculated results
underestimate the half-lives of $^{118,120,122}$Cd. This indicates
that the half-lives of an isotopic chain cannot always be reproduced
using a constant value $V_0$, and points out to a possible
isospin-dependence of $V_0$. In the present work, therefore, we
employ the following ansatz for an isospin-dependent pairing
strength:
\begin{eqnarray}\label{Eq:V0NSubZ}
  V_0 &=& V_L +\frac{V_D}{1+e^{a+b(N-Z)}}\; ,
\end{eqnarray}
and adjust the parameters to reproduce the known half-lives of even-even
nuclei with $20\leqslant Z\leqslant 50$,
and $N-Z$ in the range $8\leqslant N-Z \leqslant 36$. The resulting values:
$V_L=134.0$ MeV, $V_D=121.1$ MeV,
$a=8.5$, and $b=-0.4$ are used in the calculation
 of $\beta$-decay half-lives for
nuclei in the interval $8\leqslant N-Z \leqslant 50$.

\begin{figure*}
\begin{minipage}{0.49\linewidth}
\includegraphics[width=\columnwidth]{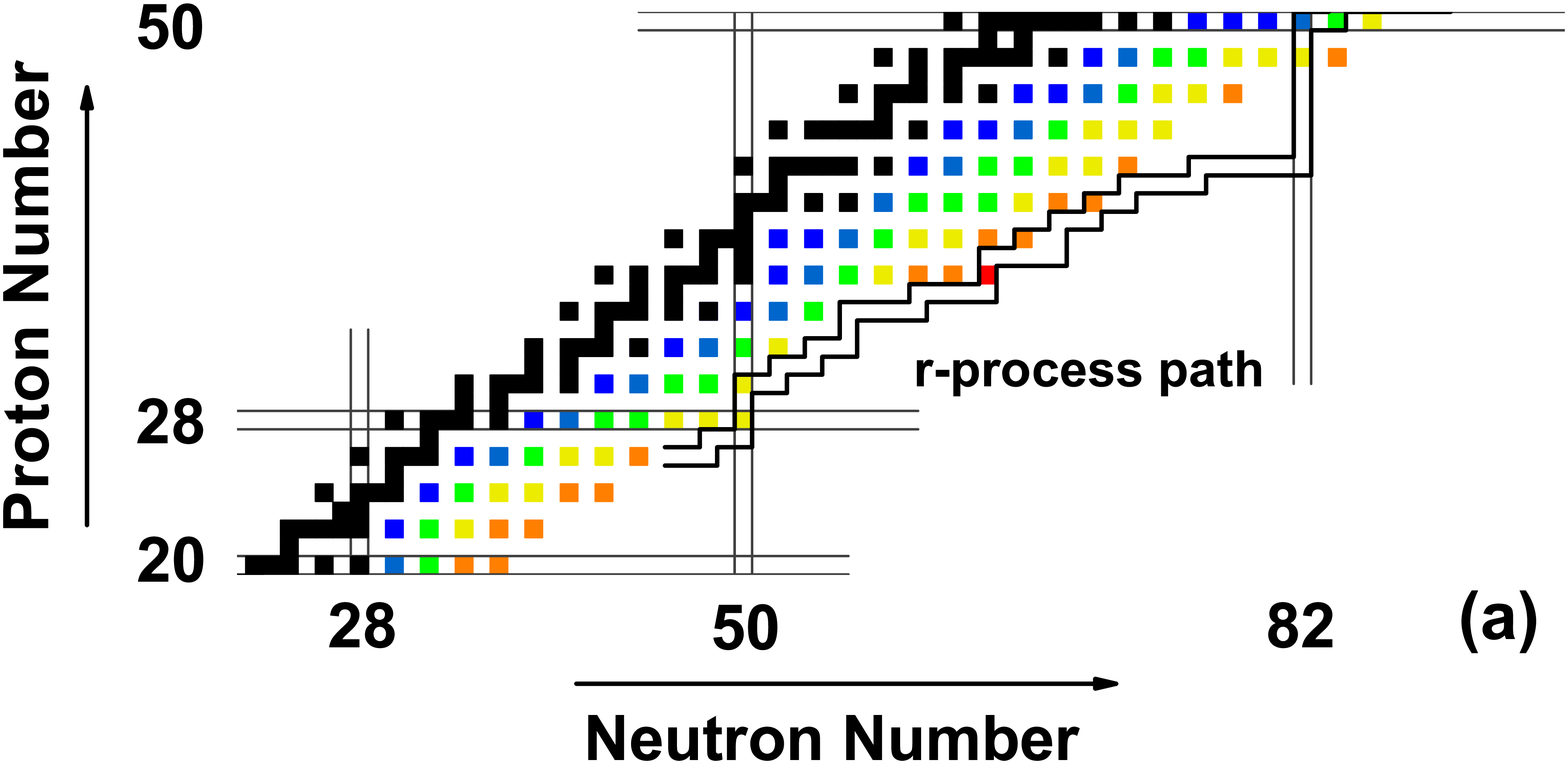}
\end{minipage}
\hfill
\begin{minipage}{0.49\linewidth}
\includegraphics[width=\columnwidth]{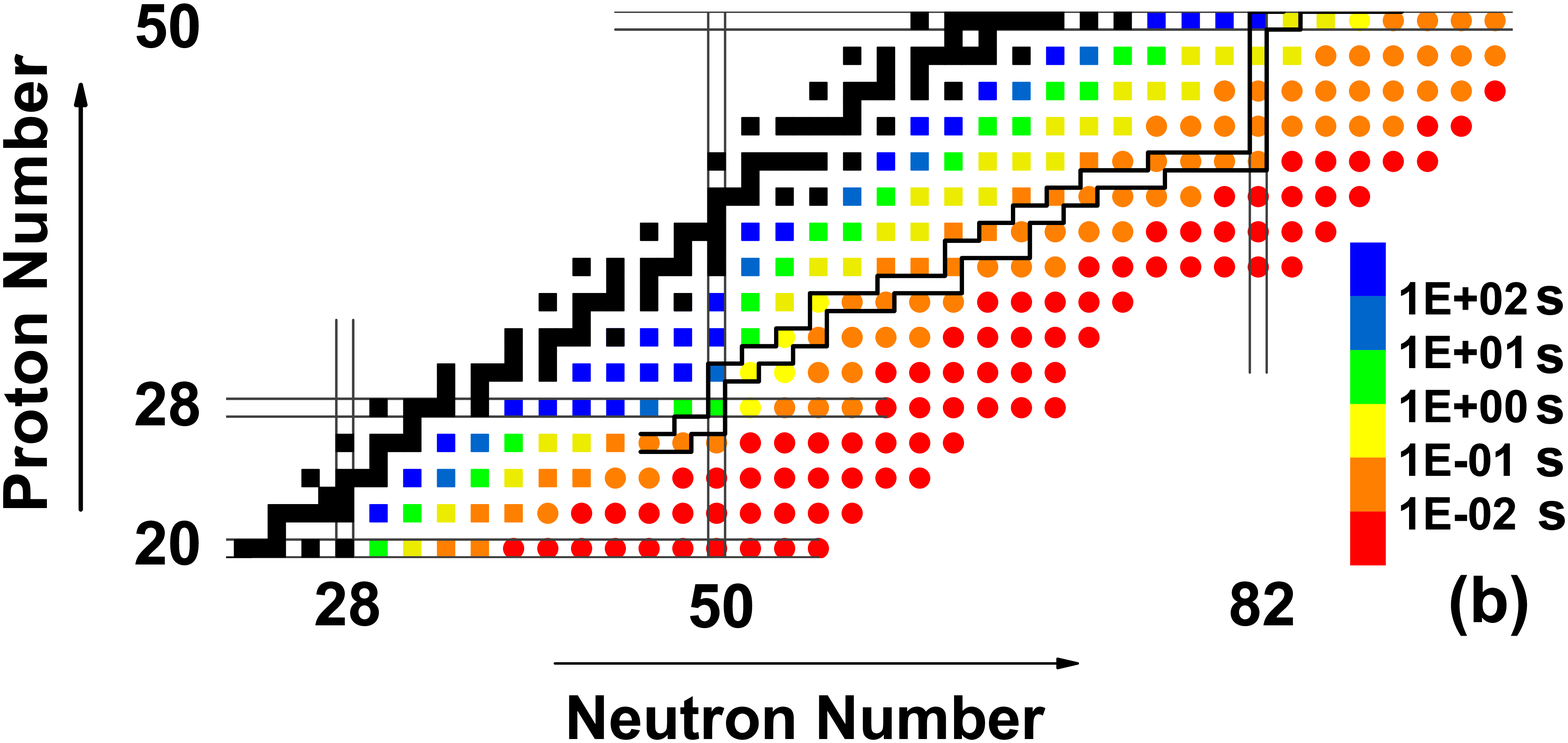}
\end{minipage}
\caption{(Color online) Contour maps of $\beta$-decay half-lives for the
$Z = 20 - 50$ even-even nuclei. The experimental
half-lives~\cite{Audi2003NPA, Nishimura2011PRL} and the
RHFB+QRPA results obtained with the effective interaction
PKO1 are shown in the right (a) and left (b) panels,
respectively. For reference, the $r$-process path calculated with the
relativistic mean-field (RMF) mass model~\cite{Geng2005PTP} is also
included in the maps.}
\label{fig2}
\end{figure*}

With the isospin-dependent strength Eq.~(\ref{Eq:V0NSubZ}) of the
proton-neutron pairing interaction, the calculated $\beta$-decay
half-lives of both the Fe and Cd isotopic chains are in excellent
agreement with data. In the next step we proceed to calculate
the half-lives of even-even nuclei with $20\leqslant Z\leqslant 50$ using
the  RHFB+QRPA model, and compare the theoretical values to data in
Fig.~\ref{fig2}. Even though a wealth of new data on nuclear $\beta$-decay
half-lives have been obtained in recent years, only few measurements can
reach the $r$-process path, especially for $r$-process nuclei around
$N=82$. The present RHFB+QRPA calculation yields results in good agreement
with the data, in particular for nuclei with half-lives $T_{1/2} <1$ s. Only
an overestimation of half-lives for Ni, Zn, and Ge isotopes near the stability
line is noticed. The longer half-lives predicted for Ni isotopes is a
common problem in self-consistent relativistic QRPA
calculations~\cite{Niksic2005PRC, Marketin2007PRC}.  For Zn and Ge
isotopes, the differences between present results and the experimental
values are remarkably reduced as the neutron number increases. Taking
$^{84}$Ge for example, the theoretical result is $1.3$ s, rather close to
the experimental value: $0.954 \pm 0.014$ s.

\begin{figure}
 \includegraphics[width=10cm]{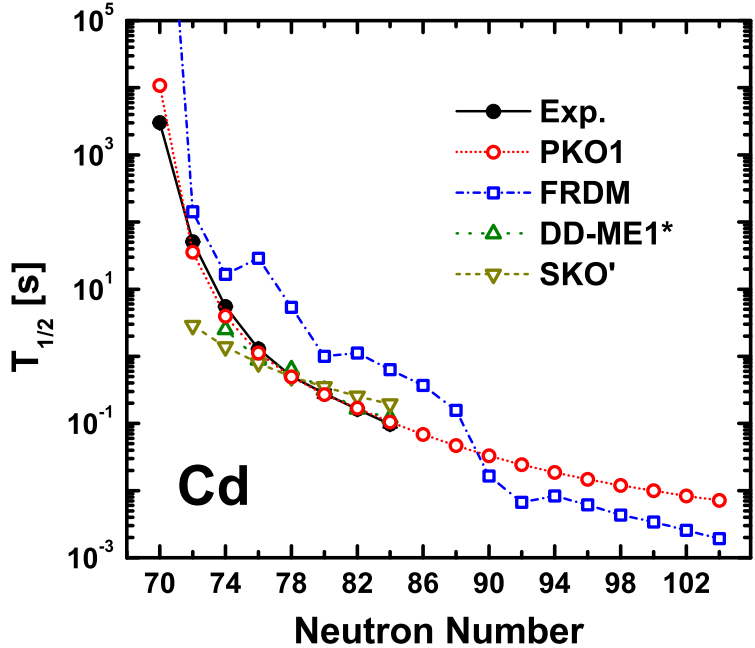}\\
 \caption{(Color online) Nuclear $\beta$-decay half-lives of Cd
isotopes calculated in the RHFB+QRPA framework using the
isospin-dependent proton-neutron pairing interaction, and the
effective interaction PKO1. For comparison, the experimental
values~\cite{Audi2003NPA}, as well as the theoretical results
obtained in the RHB+QRPA model with the effective interaction
DD-ME1$^*$~\cite{Niksic2005PRC}, the SHFB+QRPA calculation with the
effective interaction SkO$^\prime$~\cite{Engel1999PRC}, and the
FRDM+QRPA calculation~\cite{Moller1997ADNDT}, are also shown.}
\label{fig3}
\end{figure}

In Fig.~\ref{fig3}, the calculated $\beta$-decay half-lives of a sequence
of Cd isotopes using the RHFB+QRPA approach are displayed in comparison
with measured values and previous theoretical results. One notices that
the RHFB+QRPA model reproduces in detail the empirical $\beta$-decay
half-lives, whereas the RHB+QRPA \cite{Niksic2005PRC} and SHFB+QRPA
\cite{Engel1999PRC} calculations that used a constant proton-neutron
pairing strength cannot yield the appropriate isospin dependence of
$\beta$-decay half-lives. The global FRDM+QRPA calculation
\cite{Moller1997ADNDT} systematically overestimates the measured
half-lives of Cd isotopes. It has been pointed out that the overestimation
of half-lives in the FRDM+QRPA approach can at least partially be
attributed to the omission of the $T=0$ pairing~\cite{Engel1999PRC}. This
seems to be further confirmed in this work as the half-lives are
systematically reduced with the inclusion of $T=0$ pairing. For Cd
isotopes with $N\geqslant 90$, however, the FRDM+QRPA calculation yields
shorter half-lives than the RHFB+QRPA. The FRDM predicts a shape
transition from the spherical $^{136}$Cd to the deformed $^{138}$Cd with
quadrupole deformation $\beta_2=0.125$. This may indicate the shorter
half-lives for Cd isotopes with $N\geqslant 90$ obtained in the FRDM+QRPA
are due to the effect of nuclear deformation, which is not included in the
present calculation. Similar systematics are also found for the Zr, Mo,
Ru, Pd isotopes where shape transitions occur. However, it should be noted
that a recent QRPA calculation that used a Skyrme interaction has shown
the opposite trend~\cite{Sarriguren2010PRC}. Therefore, it will be
interesting to include deformation degrees of freedom into the
self-consistent QRPA calculations and study their effects on $\beta$-decay
half-lives in the future.

To analyze the impact of the predicted $\beta$-decay half-lives on
$r$-process abundances, we have also performed a classical $r$-process
calculation similar to those of Refs.~\cite{Sun2008PRC, Niu2009PRC}. In
this model, seed-nuclei (Fe) are irradiated by neutron sources of high and
continuous neutron densities $n_n$ over a timescale $\tau_r$ in a high
temperature environment ($T\sim$ 1 GK). As in Ref.~\cite{Moller2003PRC},
the components with neutron density $n_n=10^{22}-10^{24}$ cm$^{-3}$ are
used to investigate the impact of $\beta$-decay half-lives on the
$r$-matter flow. The weight $\omega$ of each $r$-process component follows
exponential relation on neutron density $n_n$:
\begin{eqnarray}
\omega(n_n)=a\times n_n^b.
\label{eq:relations}
\end{eqnarray}
The parameters $a=0.6435$ and $b=0.0411$ are taken from the calculations
in Ref.~\cite{Niu2009PRC}. The abundances for each $r$-process component
are calculated within the waiting-point approximation. In this model, the
abundance distribution in an isotopic chain is given by the Saha equation
and is entirely determined by neutron separation energies for a given
temperature $T$ (in this work, $T=1.5$ GK) and a neutron density $n_n$.
The nuclear $\beta$-decay rates control matter flow between neighboring
isotopic chains. Therefore, in this model, the $r$-process path is only
dependent on nuclear masses, from which neutron separation energies are
determined. In this work, nuclear masses are taken from the mass
evaluation~\cite{Audi2003NPA} if available, otherwise predictions of RMF
mass model~\cite{Geng2005PTP} are employed. The corresponding $r$-process
path is shown in Fig.~\ref{fig2}.

\begin{figure}
\includegraphics[width=10cm]{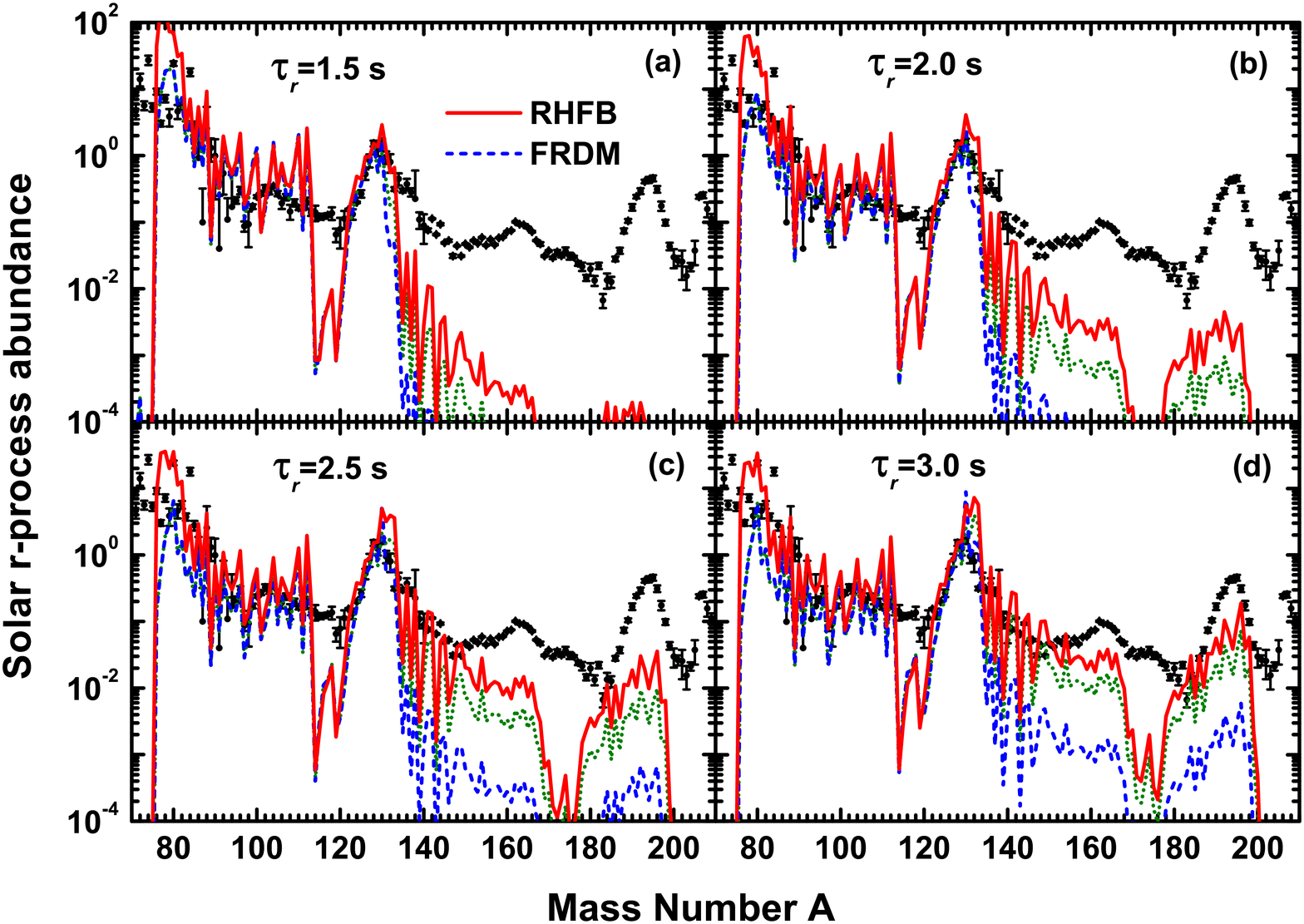}
\caption{(Color online) The impact of nuclear $\beta$-decay
half-lives on the $r$-matter flow. The solid
(dashed) curves correspond to $r$-process abundances calculated with
the RHFB+QRPA (FRDM+QRPA) $\beta$-decay half-lives in comparison to
the data denoted by the points. The dotted curves are the same as
the dashed curves but with half-lives of $^{130}$Cd and $^{134}$Sn replaced
by the corresponding RHFB+QRPA results. In all calculations, nuclear masses
are taken from the mass evaluation~\cite{Audi2003NPA} if available,
otherwise predictions of RMF mass model~\cite{Geng2005PTP} are
employed. Panels (a)-(d) correspond to the neutron irradiation times
$\tau_r = 1.5, 2.0, 2.5$, and $3.0$ s, respectively.} \label{fig4}
\end{figure}

In Fig.~\ref{fig4}, we display four snapshots of $r$-process abundances at
different neutron irradiation times $\tau_r$. One notices that the
half-lives calculated with RHFB+QRPA model (solid curves) produce a faster
$r$-matter flow in the $N=82$ region, and thus yield higher $r$-process
abundances of nuclei with $A\gtrsim 140$. To understand this difference,
the half-lives of nuclei forming the major bottlenecks of the $r$-matter
flow ($r$-process bottleneck nuclei) near $N=82$ are given in
Tab.~\ref{tbl}. Clearly, the experimental half-lives of $^{130}$Cd and
$^{134}$Sn are better reproduced by RHFB+QRPA approach, while the
FRDM+QRPA approach remarkably overestimates these half-lives. By merely
replacing half-lives of $^{130}$Cd and $^{134}$Sn in the FRDM+QRPA
approach by the corresponding RHFB+QRPA results, the calculated
$r$-process abundances are shown by the dotted curves in Fig.~\ref{fig4}.
It is clear that the differences of the calculated abundances are mainly
due to the differences of predicted half-lives of $^{130}$Cd and
$^{134}$Sn between RHFB+QRPA and FRDM+QRPA approach. Furthermore, if we
replace the FRDM+QRPA half-lives of all nuclei in Tab.~\ref{tbl} by the
corresponding RHFB+QRPA results, almost the same abundances are obtained
for nuclei around and beyond the second abundance peak. Moreover, by
summing up the half-lives of $r$-process bottleneck nuclei near $N=82$,
one can roughly estimate the time when the $r$-process passes the second
abundance peak. Based on the RHFB+QRPA results this time is speeded up to
$1.00$ s, from the $4.83$ s predicted by the FRDM+QRPA calculation. This
is an important result for the estimate of the duration of the
$r$-process, and hence the origin of heavy elements in the universe. In
addition, using the RHFB+QRPA results, higher abundances are found at
$A\sim 80$. This can easily be understood because it takes more time to
pass the $r$-path nuclei $^{78}$Ni and $^{80}$Zn and, as a result, higher
abundances are accumulated.

\begin{table}
\begin{center}
\caption{The half-lives of $r$-process bottleneck nuclei near $N=82$
predicted by RHFB+QRPA and FRDM+QRPA approaches. For comparison, the
experimental values~\cite{Audi2003NPA} are also shown if available.} \label{tbl}
\begin{tabular}{ccccc}
\hline \hline
Nucleus      &~~ &\multicolumn{3}{c}{Half-life (s)}                  \\
\cline{3-5}
             &~~ &RHFB+QRPA        &FRDM+QRPA    &Exp.               \\
\hline
$^{124}$Mo   &~~ &0.0108           &0.0106       &---                \\
$^{126}$Ru   &~~ &0.0205           &0.0342       &---                \\
$^{128}$Pd   &~~ &0.0486           &0.1251       &---                \\
$^{130}$Cd   &~~ &0.1685           &1.1232       &$0.162 \pm 0.007$  \\
$^{134}$Sn   &~~ &0.7530           &3.5410       &$1.050 \pm 0.011$  \\
\hline \hline
\end{tabular}
\end{center}
\end{table}



In summary, we have introduced a fully self-consistent proton-neutron
quasiparticle random phase approximation (QRPA) for spherical nuclei,
based on the relativistic Hartree-Fock-Bogoliubov (RHFB) framework. By
employing an isospin-dependent proton-neutron $T=0$ pairing interaction,
the RHFB+QRPA model has been applied to a study of  $\beta$-decay
half-lives of neutron-rich nuclei with $20\leqslant Z\leqslant 50$,
extending over the entire $r$-process path from $N = 50$ to $N = 82$. It
has been found that the RHFB+QRPA model calculation reproduces the
experimental $\beta$-decay half-lives for neutron-rich nuclei, especially
for nuclei with half-lives less than one second. Using the calculated
$\beta$-decay half-lives of neutron-rich nuclei a remarkable speeding up
of $r$-matter flow has been predicted. This leads to an enhancement of
$r$-process abundances of elements with $A\gtrsim 140$. It should be
pointed out that the speeding up of $r$-matter flow mainly results from
shorter half-lives of $r$-path nuclei close to $N=82$, which are spherical
or nearly spherical. Therefore, the spherical approximation in this work
does not influence our key conclusions. Of course, to quantitatively
reproduce the $r$-process abundances in the whole region the deformed QRPA
approach should be employed, and we leave this further refinement for
future consideration.

\bigskip
\bigskip


This work was partly supported by the Major State 973 Program
2013CB834400, the National Natural Science Foundation of China under Grant
Nos. 10975008, 11105006, 11175001, 11175002, 11075066, and 11205004, the
211 Project of Anhui University under Grant No. 02303319-33190135, the
Fundamental Research Funds for Central Universities under Contract No.
lzujbky-2012-k07, the Program for New Century Excellent Talents in
University of China under Grant No. NCET-10-0466, the Grant-in-Aid for
JSPS Fellows under Grant No. 24-02201, and the MZOS-project No.
1191005-1010.




\end{document}